\documentclass[11pt]{article}
\usepackage[margin=1in]{geometry}
\usepackage{setspace}
\usepackage{titlesec}
\usepackage{natbib}
\usepackage{subcaption}
\usepackage{placeins}
\usepackage{amsmath, amssymb}
\usepackage{graphicx}
\usepackage{booktabs}
\usepackage{caption}
\usepackage{enumitem}
\usepackage[colorlinks=true, citecolor=blue, urlcolor=blue]{hyperref}
\usepackage{url}
\usepackage{multirow}
\usepackage{xcolor}
\usepackage{fancyhdr}
\titleformat{\section}{\normalfont\large\bfseries}{\thesection.}{1em}{}
\titleformat{\subsection}{\normalfont\normalsize\bfseries}{\thesubsection.}{1em}{}
\title{Causal Analysis of Health, Education, and Economic Well-Being \\ in India - Evidence from the Young Lives Survey}
\author{
 Anushka De$^1$ ,  Diganta Mukherjee$^1$\thanks{We thank Prof. Ravi Kanbur for his comments and suggestions on earlier stages of this work. The usual caveat applies.}\\
 $^1$Indian Statistical Institute, Kolkata\\
anushka.isical@gmail.com, digantam@hotmail.com
}

\date{}
\begin{document}
\maketitle
\begin{abstract}
This study investigates the dynamic and potentially causal relationships among childhood health, education, and long-term economic well-being in India using longitudinal data from the Young Lives Survey. While prior research often examines these domains in isolation, we adopt an integrated empirical framework combining panel data methods, instrumental variable regression, and causal graph analysis to disentangle their interdependencies. Our analysis spans five survey rounds covering two cohorts of children tracked from early childhood to young adulthood. Results indicate strong persistence in household economic status, highlighting limited intergenerational mobility. Education, proxied by Item Response Theory-based mathematics scores, consistently emerges as the most robust predictor of future economic well-being, particularly in the younger cohort. In contrast, self-reported childhood health shows limited direct impact on either education or later wealth, though it is influenced by household economic conditions. These findings underscore the foundational role of wealth and the growing importance of cognitive achievement in shaping life trajectories. The study supports policy approaches that prioritize early investments in learning outcomes alongside targeted economic support for disadvantaged households. By integrating statistical modeling with development policy insights, this research contributes to understanding how early-life conditions shape economic opportunity in low- and middle-income contexts.

\textbf{Keywords}: Causal analysis, Education, Economic Mobility, Longitudinal data, Subjective well-being 
\end{abstract}

\section{Introduction}
The role of early-life conditions in shaping long-term human development outcomes has been a central concern in both economic policy and statistical research. In particular, childhood health and education are widely recognized as foundational components of human capital, influencing future earnings, employment prospects, and broader measures of well-being. 

The relationship between childhood health and later educational outcomes has been well-established across both developed and developing countries. In Canada, \citet{Currie_Stabile} demonstrate that children in poorer health are more likely to repeat grades and perform worse in school. In developing country contexts, \citet{Currie_Vogal_2013} highlight that health shocks during childhood can significantly influence educational attainment, economic productivity, and overall well-being in later life.
Similarly, cognitive ability as a key outcome of educational investment is consistently associated with improved economic outcomes. \citet{ROHDE2007}, in a meta-analysis of studies from high-income countries, demonstrate that cognitive skills are strong predictors of academic achievement. \citet{hanushek2008role} extend this argument globally, showing that cross-country differences in cognitive skills significantly explain variations in national GDP growth. Additional evidence from \citet{Ozawa_Review_LIM_2022} underscores that in developing regions such as Sub-Saharan Africa, Latin America, and South Asia, higher cognitive ability is linked to improved schooling and employment outcomes, highlighting the need for context-sensitive policy design.
The long-term economic penalties of poor childhood health are also well-documented. Using British cohort data, \citet{CASE2005365} find that individuals with adverse childhood health  such as low birth weight or chronic conditions experience lower educational attainment, reduced labor force participation, and diminished occupational status. \citet{Smith2009} provides similar evidence from the U.S., showing that poor childhood health leads to significantly worse adult labor market outcomes, independent of parental background. In developing countries, the impact may be even more severe due to weaker health infrastructure. For example, \citet{victora2008maternal} link childhood malnutrition in Brazil to reduced wages and productivity in adulthood, particularly among poorer households. Collectively, these studies highlight how early health disadvantages perpetuate inequality across generations, especially in resource constrained settings like India.

Economic mobility, the extent to which individuals or families can move up or down the economic ladder across generations, is a key indicator of inequality and opportunity in a society. \citet{BlackDevereux2011} provide a comprehensive synthesis of intergenerational mobility studies, highlighting how economic status is often transmitted from parents to children, with limited upward movement in low-mobility contexts. \citet{Corak2013} further shows that countries with higher income inequality tend to exhibit lower intergenerational mobility, a pattern that underscores the deep interlinkages between inequality and opportunity. In the Indian context, where reliable income data are often scarce and noisy, \citet{EconomicMobility_India} develop a partial identification approach to estimate bounds on mobility using survey data. Their findings suggest that conventional point estimates may understate both upward and downward mobility, emphasizing the importance of robust statistical frameworks when analyzing persistence in economic well-being. These insights are particularly relevant in settings like India, where early-life disadvantages, if left unaddressed, may perpetuate long-term economic inequality.

India presents a particularly compelling context for studying the long-term effects of childhood health and education on economic outcomes. As one of the fastest-growing major economies, it continues to grapple with deep-rooted disparities in access to healthcare, education, and wealth, especially across caste, class, and regional lines. The coexistence of rapid economic development with persistent structural inequality makes it critical to understand how early-life disadvantages shape intergenerational mobility. Moreover, India’s large-scale policy interventions in child health and education over the past two decades provide a relevant backdrop for identifying which pathways may yield the greatest returns in terms of long-term economic well-being. 

Despite extensive research in each of these domains, few studies have jointly examined their causal interconnections, particularly in the context of developing economies such as India.

While previous research has examined the isolated effects of health or education on adult economic outcomes, the potentially causal interrelationships among these domains remain underexplored, particularly in developing economies. Evidence from high-income contexts, such as \citet{Bellani2018}'s analysis of European Union countries, highlights education’s mediating role in linking childhood poverty to adult income. However, in countries like India  where social mobility is constrained by structural inequalities, a dynamic understanding of how health, education, and wealth interact is critical for evidence-based policy design.

This paper addresses this gap by first analyzing the persistence of economic well-being and then evaluating the interrelationships among childhood health, education, and economic well-being of the family. While \citet{Bellani2018} employ a causal mediation framework, this study adopts a complementary approach using panel data methods, instrumental variable regression, and causal graph analysis to disentangle the temporal and reciprocal linkages among these factors during a critical developmental period. These techniques enable the identification of statistically robust pathways, offering insight into the most effective domains for policy intervention. 

This study makes three distinct contributions to the literature on intergenerational development and applied statistics in low- and middle-income countries.

First, it provides one of the few empirical analyses that jointly models the causal pathways linking childhood health, education, and economic well-being using longitudinal data from a large developing country. While prior studies often examine these domains in isolation, our study leverages panel data and identification strategies to disentangle their dynamic interrelationships in the Indian context. 
Second, this paper offers a methodologically integrated framework, combining instrumental variable techniques, panel regression, and causal graph analysis to navigate challenges of endogeneity, temporal structure, and reciprocal causality. By applying this approach to the Young Lives India dataset, it demonstrates how causal inference methods can be adapted for development-focused social statistics with data limitations such as self-reported measures and asset-based proxies.
Third, the study generates policy-relevant insights by identifying cognitive achievement as a stronger and more consistent predictor of future economic well-being than childhood health, after adjusting for wealth persistence. This highlights the growing importance of education-focused investments in promoting upward mobility, particularly for children in lower-income settings, and reinforces calls for integrated early-life policy interventions.
Collectively, this work contributes to the applied statistics literature by demonstrating how robust causal inference methods can be deployed to inform development policy, particularly in countries where structural inequality, data quality issues, and policy relevance must be addressed jointly.

This study draws on data from the Young Lives Survey \citep{Young_Lives_Oxford}, a longitudinal study of childhood poverty conducted in four countries: Ethiopia, India, Peru, and Vietnam. In India, the survey has tracked approximately 3,000 children from Andhra Pradesh and Telangana since 2002, across two cohorts born seven years apart. The strength of the Young Lives dataset lies in its ability to trace long-term changes in child, household, and community-level indicators, offering a unique opportunity to examine developmental trajectories and their socioeconomic consequences. 

The remainder of the paper is organized as follows. Section~\ref{sec:Data} describes the dataset in detail, along with key variables and preliminary descriptive statistics. Section~\ref{sec:methodology} outlines the analytical strategy, including model specifications and identification approaches. Section~\ref{sec:results} presents the main empirical findings. Section~\ref{sec:conclusion} summarizes the results and acknowledges key limitations. Finally, Section~\ref{sec:discussion} reflects on broader policy implications and directions for future research.

\section{Data}
\label{sec:Data}
\subsection{Young Lives Survey}

The Young Lives Survey is a longitudinal study designed to investigate the dynamics of childhood poverty across four low- and middle-income countries: Ethiopia, India, Peru, and Vietnam. The study tracks 12,000 children over a 20-year period, spanning five in-person survey rounds between 2002 and 2016, followed by a sixth phone-based round conducted during the COVID-19 pandemic in 2020–2021.

Young Lives is unique in the Indian context as the only longitudinal dataset that adopts a life-course perspective, covering a wide array of domains including nutrition, health, education, psychosocial development, and transitions to the labour market. Two cohorts of children were enrolled in 2002: a Younger Cohort aged approximately one year, and an Older Cohort aged around eight years. The cohorts were subsequently surveyed at ages 5 and 12 (Round 2), 8 and 15 (Round 3), 12 and 19 (Round 4), and 15 and 22 (Round 5), respectively. This design allows for the tracking of developmental trajectories from infancy through adolescence and early adulthood.

For this study, we focus on Rounds 1 to 5 to ensure consistency in data collection methods and variable definitions \citep{Boyden2022_data}. Round 6, conducted via phone during the pandemic, introduces substantial differences in design and content and is therefore excluded. Future research may incorporate this round to explore the effects of COVID-19 and methodological adaptations in longitudinal survey research.

\subsection{Variables and Measurement}
This study focuses on three core dimensions of child development captured in the Young Lives dataset - health, education, and economic well-being, each represented by proxy variables selected for consistency, availability across rounds, and relevance to long-term outcomes. Detailed information on the construction of variables, sampling strategy and descriptive summaries for the Indian dataset are available in \citet{Singh_India_report_YoungLives}.

Childhood health is proxied using the child’s self-reported subjective well-being, recorded on a 9-point ordinal scale, where higher values indicate better perceived health status. Subjective health assessments are widely used in empirical research, as they provide a holistic view of perceived well-being and have been shown to correlate with objective health outcomes \citep{Currie_Stabile}. This variable is available for the Older Cohort beginning in Round 2 and for the Younger Cohort from Round 3 onward.

Educational attainment is measured using test scores derived through Item Response Theory \citep{chen_IRT}, administered consistently across survey rounds. This study focuses specifically on mathematics scores to represent cognitive achievement, given their longitudinal consistency and demonstrated relevance as predictors of labor market outcomes \citep{hanushek2008role}. Math IRT scores are available for the Older Cohort in Rounds 2, 3, and 4, and for the Younger Cohort beginning in Round 3.

Economic well-being is measured using a composite wealth index, constructed from household ownership of durable assets and access to basic services. In settings where reliable income data are unavailable, asset-based indices are widely accepted as proxies for long-run economic status \citep{filmer2001estimating}. Wealth index data are available across all five rounds for both cohorts.

In addition to the three primary variables, mother’s education is included as a control variable in some of the models (Section \ref{sec: health ->edu}). It is measured as the number of years of completed schooling and serves as a proxy for household-level educational background, which may influence both child health and learning outcomes.

\subsection{Data Processing}
The dataset was already structured in panel format, with repeated observations for each child across multiple survey rounds. For this study, we retained only those observations where all three key variables- subjective well-being (health), mathematics IRT scores (education), and the household wealth index (economic well-being), were jointly available in the relevant rounds. All data processing and statistical analyses were performed using R version 4.1.1.

To address sparsity in the tails of the subjective well-being scale, levels 1 and 2 were combined into a single lower category, and levels 7 through 9 into a single upper category. This recoding improved the reliability of statistical estimation by ensuring sufficient representation across ordinal levels.

\subsection{Summary Statistics and Sample Description}
The analytical sample includes children from both the Younger and Older Cohorts, observed across Rounds 1 to 5. Due to the availability of math test scores and subjective well-being, some rounds are excluded for certain analyses (e.g., Rounds 1 \& 2 for the Younger Cohort). After restricting to non-missing observations, the final sample consists of approximately 9,510 observations for health, 8,480 for education, and 14,556 for economic well-being, aggregated across rounds and cohorts. 

Table \ref{tab:desc_stats} summarizes the distribution of the three core variables: subjective well-being (health), mathematics IRT scores (education), and the wealth index (economic well-being), by survey round and cohort.
Table \ref{tab:pairwise_correlation} presents the pairwise correlations between each combination of these variables. For the association between wealth index and math IRT scores, Pearson's correlation coefficient is reported, as both variables are continuous. In contrast, Spearman's rank correlation is used to assess relationships involving subjective well-being, which is an ordinal categorical variable. This approach accounts for the non-linear and ordinal nature of the health measure while enabling comparison across domains.
\begin{table}[h!]
\caption{Descriptive statistics for health, education, and economic well-being by cohort and survey round}
\label{tab:desc_stats}
\centering
\begin{tabular}{lrl lrll}
\hline
\multirow{2}{*}{\textbf{Variables}}    & \multicolumn{3}{c}{\textbf{Older   Cohort}}                                      & \multicolumn{3}{c}{\textbf{Younger   Cohort}}                                             \\ \cline{2-7}  & \multicolumn{1}{l}{\textbf{Round}}   & \multicolumn{1}{l}{\textbf{Mean}}   & \textbf{SD} & \multicolumn{1}{l}{\textbf{Round}}   & \multicolumn{1}{l}{\textbf{Mean}}   & \textbf{SD} \\ \hline
\multirow{4}{*}{Subjective Well-Being} & \multicolumn{1}{c}{2} & \multicolumn{1}{l}{2.68}   & 1.44                 & \multicolumn{1}{c}{3} & \multicolumn{1}{l}{3.81}   & 1.69                 \\  & \multicolumn{1}{c}{3} & \multicolumn{1}{l}{3.66}   & 1.55                 & \multicolumn{1}{c}{4} & \multicolumn{1}{l}{3.55}   & 1.35                 \\   & \multicolumn{1}{c}{4} & \multicolumn{1}{l}{3.91}   & 1.3                  & \multicolumn{1}{c}{5} & \multicolumn{1}{l}{3.99}   & 1.25 \\   & \multicolumn{1}{c}{5} & \multicolumn{1}{l}{3.95}   & 1.29                 & \multicolumn{3}{l}{}    \\ \hline
\multirow{3}{*}{Maths IRT Scores}      & \multicolumn{1}{c}{2} & \multicolumn{1}{l}{501.97} & 97.77                & \multicolumn{1}{c}{3} & \multicolumn{1}{l}{347.95} & 79.29    \\   & \multicolumn{1}{c}{3} & \multicolumn{1}{l}{467.34} & 93.48                & \multicolumn{1}{c}{4} & \multicolumn{1}{l}{463.98} & 87.66 \\  & \multicolumn{1}{c}{4} & \multicolumn{1}{l}{494.23} & 111.71               & \multicolumn{1}{c}{5} & \multicolumn{1}{l}{491}    & 91.98                \\ \hline
\multirow{5}{*}{Wealth Index}          & \multicolumn{1}{c}{1} & \multicolumn{1}{l}{0.41}   & 0.21                 & \multicolumn{1}{c}{1} & \multicolumn{1}{l}{0.41}   & 0.2                  \\    & \multicolumn{1}{c}{2} & \multicolumn{1}{l}{0.47}   & 0.2                  & \multicolumn{1}{c}{2} & \multicolumn{1}{l}{0.46}   & 0.2                  \\   & \multicolumn{1}{c}{3} & \multicolumn{1}{l}{0.52}   & 0.17                 & \multicolumn{1}{c}{3} & \multicolumn{1}{l}{0.51}   & 0.18                 \\   & \multicolumn{1}{c}{4} & \multicolumn{1}{l}{0.61}   & 0.15                 & \multicolumn{1}{c}{4} & \multicolumn{1}{l}{0.59}   & 0.17                 \\   & \multicolumn{1}{c}{5} & \multicolumn{1}{l}{0.65}   & 0.15                 & \multicolumn{1}{c}{5} & \multicolumn{1}{l}{0.63}   & 0.16                 \\ \hline
\end{tabular}
\vspace{0.5em}\\
{\footnotesize $\dagger$ SD = Standard Deviation. All statistics are pooled across cohorts and rounds.}
\end{table}

\begin{table}[h!]
\caption{ Pairwise correlations between subjective well-being (health), mathematics IRT scores (education), and wealth index (economic well-being)}
\label{tab:pairwise_correlation}
\centering
\begin{tabular}{lllll}
\hline
\multirow{2}{*}{\textbf{Variables}}                        & \multicolumn{2}{l}{\textbf{Older   Cohort}} & \multicolumn{2}{l}{\textbf{ Younger   Cohort}} \\ \cline{2-5}  & \textbf{Round}         & $r$         & \textbf{ Round}          & $ r$          \\ \hline
\multirow{3}{*}{\textbf{Health \& Education}}              & \multicolumn{1}{c}{2}                      & 0.033              & \multicolumn{1}{c}{3}                       & 0.012               \\  & \multicolumn{1}{c}{3}                     & 0.046              & \multicolumn{1}{c}{4}                       & 0.05                \\ 
                     & \multicolumn{1}{c}{4}                      & 0.075              & \multicolumn{1}{c}{5}                       & 0.065               \\ \hline
\multirow{3}{*}{\textbf{Education \& Economic Well-Being}} & \multicolumn{1}{c}{2}                      & 0.241    & \multicolumn{1}{c}{3}                       & 0.292               \\ 
  & \multicolumn{1}{c}{3}                  & 0.316              & \multicolumn{1}{c}{4}                       & 0.332               \\ 
     & \multicolumn{1}{c}{4}                   & 0.293              & \multicolumn{1}{c}{5}                     & 0.365               \\ \hline
\multirow{4}{*}{\textbf{Health \& Economic Well-Being}}    & \multicolumn{1}{c}{2}                   & 0.102              & \multicolumn{1}{c}{3}                        & 0.07                \\ 
      & \multicolumn{1}{c}{3}                 & 0.08               & \multicolumn{1}{c}{4}                      & 0.085               \\  
 & \multicolumn{1}{c}{4}                   & 0.063              & \multicolumn{1}{c}{5}                    & 0.072               \\ 
     & \multicolumn{1}{c}{5}                    & 0.085              & \multicolumn{2}{l}{}                          \\ \hline
\end{tabular}
\vspace{0.5em}\\
{\footnotesize $\dagger$ $r$ is Pearson's correlation is for continuous–continuous pairs (education, economic wealth index), and Spearman's rank correlation when an ordinal variable (subjective well-being) is involved. }
\end{table}
The correlations in Table \ref{tab:pairwise_correlation} reveal positive associations between each pair of variables. As expected, higher subjective well-being tends to be associated with both better educational outcomes (math scores) and higher economic status (wealth index). Similarly, education and wealth are positively correlated, reflecting established links between cognitive achievement and household economic status. While the correlation coefficients provide useful information about the pairwise associations between health, education, and economic well-being, they do not imply causal relationships. Correlation can arise due to a variety of reasons including reverse causality, omitted variable bias, or common underlying factors and thus does not establish direction or mechanism. As emphasized by \citet{pearl2009causality}, causal inference requires a model of the data-generating process and appropriate statistical tools to identify causal effects. Similarly, \citet{angrist2009mostly} argue that econometric techniques like instrumental variables and panel data models are essential to move from descriptive correlations toward credible causal claims. 
In the following sections, we apply such methods to uncover the directional and potentially causal linkages among these variables.

\section{Methodological Framework}
\label{sec:methodology}
This section outlines the modeling framework used to analyse the directional and potentially causal relationships between childhood health, education, and economic well-being. Given the longitudinal structure of the data and the possibility of endogeneity, the study employs a combination of regression models, instrumental variable (IV) techniques, and causal graph analysis to disentangle interdependencies and identify robust predictors of long-term outcomes. All regression models are estimated using heteroskedasticity-robust (White) standard errors \citep{White} to ensure valid inference in the presence of non-constant residual variance.

\subsection{Modeling the Persistence of Economic Well-being}
\label{sec:persistence_model}
To investigate the temporal stability of household economic status, we begin by modeling the persistence of economic well-being using the wealth index across survey rounds. Specifically, we estimate a linear regression of the current wealth index on its past value within each cohort:

\begin{equation}
    \text{Wealth}_{it} = \alpha + \beta \cdot \text{Wealth}_{i,t-1} + \varepsilon_{it}
    \label{equ:persistence model}
\end{equation}

where $\text{Wealth}_{it}$ denotes the wealth index of individual $i$ in round $t$, and $\text{Wealth}_{i,t-1}$ represents the individual's wealth index in the previous round. 
The coefficient $\beta$ captures the extent of persistence; a value close to 1 indicates high temporal stability, while a value closer to 0 suggests economic mobility.
This regression is estimated separately for each cohort-round pair where both current and lagged wealth data are available. The $R^2$ values are also reported to quantify the explanatory power of past economic status (see Table \ref{tab:results_persistence_model}). 
As discussed before, all regressions are estimated using robust standard errors to ensure valid inference. 

The estimated degree of persistence, combined with residual variation that remains unexplained, motivates the inclusion of additional predictors such as past health and education in subsequent models to further account for variation in economic outcomes.
\subsection{Analyzing the Influence of Health on Education}
\label{sec: health ->edu}
This subsection investigates whether a child’s subjective health status influences their educational achievement, as measured by math IRT scores. We begin with a baseline linear regression of education on health:

\begin{equation}
    \text{Education}_{it} = \alpha + \gamma \cdot \text{Health}_{it} + \varepsilon_{it}
    \label{equ:direc edu =health model}
\end{equation}

where $\text{Education}_{it}$ denotes the math IRT score of individual $i$ at round $t$, and $\text{Health}_{it}$ is the individual's self-reported subjective well-being in the same round. The coefficient $\gamma$ captures the contemporaneous association between health and cognitive performance.

However, health may be an endogenous regressor due to reverse causality or unobserved confounding factors, such as household environment, that simultaneously affect both health and education.
To formally assess endogeneity, the Durbin–Wu–Hausman test \citep{durbin1954errors, wu1973note, hausman1978specification} is employed. This test evaluates whether an explanatory variable is endogenous by comparing the consistency of ordinary least squares (OLS) and instrumental variable (IV) estimates. The instrumental variable method uses external variables known as instruments that are correlated with the endogenous regressor but uncorrelated with the error term, allowing for consistent estimation of causal effects. The previous round’s health value serves as an instrument for current health, satisfying both these conditions.

To address potential endogeneity, we estimate a two-stage least squares (2SLS) model. In the first stage, current-round health is regressed on lagged health, current wealth index, and lagged wealth index (see table \ref{tab:results_first_stage}). In the second stage, current IRT scores (education) are regressed on the residual values from the first stage (actual health -predicted health), along with mother’s education, and both current and present wealth index values (table \ref{tab:results_second_stage}). This approach helps isolate the variation in health that is exogenous to unobserved confounders, thereby allowing for a more credible estimate of its causal effect on education. 

In the first stage, the response variable - subjective well-being, is ordinal in nature, making ordinary least squares inappropriate. Therefore, a cumulative link model is employed to account for the ordered categorical structure of the outcome \citep{Agresti}. In the second stage, since the IRT scores (education), the fitted values from the first stage, and the wealth index are all continuous, ordinary least squares regression is suitable. Mother’s education, although originally coded in categorical levels, spans enough gradations to be treated as a continuous covariate in the analysis.
To assess the statistical significance of the predictors in the first-stage cumulative link model, the Wald test is used, where the test statistic is calculated as the square of the estimated coefficient divided by its estimated variance. 
This statistic follows a chi-square distribution with degrees of freedom equal to the number of parameters tested. A small p-value (typically $< 0.05$) indicates that the predictor significantly contributes to explaining variation in the ordinal response.

In the second-stage OLS regression, standard $t$-statistics are used to test the significance of individual coefficients. In both stages, robust standard errors are employed to ensure valid inference in the presence of non-constant residual variance.

This framework allows us to assess whether improvements in self-reported health are causally linked to higher cognitive achievement, beyond mere correlation. Estimation results are presented in Section 4.
\subsection{Analyzing the Influence of Health and Education on Economic Well-being}
\label{sec:health_edu_wealth}

This subsection investigates the joint influence of early-life health and cognitive ability on subsequent economic well-being, as measured by the current round's household wealth index.

To isolate the contribution of health and education beyond the persistence of wealth itself, the persistence model described in Section~\ref{sec:persistence_model} is estimated (table \ref{tab:results_cohort_persistence}). The residuals from that model, denoted as $\text{Wealth}_{r,it}$, capture the unexplained component of current wealth not accounted for by its lagged value. These residuals are then used as the response variable in a regression on past subjective well-being and educational scores. By regressing on residual wealth, we ensure that the estimated influence of health and education is orthogonal to persistence, thereby isolating their incremental explanatory power.

The modeling framework is specified as:
\begin{equation}
    \text{Wealth}_{r,it} = \alpha + \beta_1 \cdot \text{Health}_{i,t-1} + \beta_2 \cdot \text{Education}_{i,t-1}  + \varepsilon_{it}
    \label{equ:past_health_edu_wealth}
\end{equation}

where $\text{Wealth}_{r,it}$ is the residual wealth index for individual $i$ in round $t$, and $\text{Health}_{i,t-1}$ and $\text{Education}_{i,t-1}$ represent the previous round's subjective well-being and IRT-based test scores, respectively.
Model~\ref{equ:past_health_edu_wealth} is estimated using ordinary least squares (OLS) with robust standard errors to ensure valid inference. The statistical significance of individual coefficients is assessed using standard $t$-statistics, and model fit is evaluated using the $R^2$ statistic (Tables \ref{tab:results_wealth_residual_model}, \ref{tab: results_ Adjusted R^2}).

This specification allows us to examine whether improvements in childhood health and educational achievement contribute to better economic well-being in subsequent rounds, beyond what is explained by past wealth alone.
\subsection{Causal Graphs}
\label{sec:causal_graphs}
This section synthesizes the regression-based findings from Sections \ref{sec: health ->edu} and \ref{sec:health_edu_wealth} through a graphical analysis of the potential causal relationships among health, education, and economic well-being. Causal graphs, or Directed Acyclic Graphs (DAGs), offer an intuitive and rigorous framework for visualizing assumptions about the directionality and structure of dependencies among variables \citep{pearl2009causality}.

Edges in the graph represent statistically significant direct effects, while the absence of an edge implies conditional independence under the assumed model. These graphical representations facilitate the identification of potential confounding variables, mediators, and backdoor paths that must be addressed in causal inference.

The DAGs constructed in this study serve three main purposes: (i) to summarize relationships supported by both empirical results and theoretical understanding, (ii) to clarify the conditional independencies underlying the identification strategy, and (iii) to guide future extensions, including simulation or intervention-based policy analysis. We discuss in detail in section \ref{sec:results_causal_graphs}.


\section{Results}
\label{sec:results}

This section presents the main empirical findings from the models discussed in Section~\ref{sec:methodology}. The results are organized according to the specific research questions addressed: the persistence of economic well-being, the causal influence of health on education, and the joint effects of health and education on economic outcomes. 

\subsection{Persistence of Economic Well-being}
\label{sec:results_persistence}
Table~\ref{tab:results_persistence_model} shows that the wealth index exhibits strong temporal persistence across both cohorts.
\begin{table}[h!]
\caption{Results from Model \ref{equ:persistence model}}
\label{tab:results_persistence_model}
\centering
\begin{tabular}{lllllll}
\hline
\multicolumn{1}{c}{\multirow{2}{*}{\textbf{Term}}} & \multicolumn{3}{c}{\textbf{Older   Cohort}}                                         & \multicolumn{3}{c}{\textbf{Younger   Cohort}}                                       \\ \cline{2-7} 
\multicolumn{1}{c}{}                                    & \multicolumn{1}{c}{\textbf{Estimate}} & \multicolumn{1}{c}{\textbf{SD}} & \multicolumn{1}{c}{\textbf{\textit{p-value}}} & \multicolumn{1}{c}{\textbf{ Estimate}} & \multicolumn{1}{c}{\textbf{SD}} & \multicolumn{1}{c}{\textbf{\textit{p-value}}} \\ \hline
Intercept                                              & 0.24   & 0.006                  & $<10^{-16}$                 & {  0.22}                         & 0.005                  & $<10^{-16}$                 \\ 
$W_{t-1}$                                                 & 0.54                         & 0.014       & $<10^{-16}$  & {  0.58}                         & 0.01                   & $<10^{-16}$                 \\ \hline
\end{tabular}
\vspace{0.5em}\\
\begin{flushleft}
{\footnotesize $\dagger$ SD = Robust Standard Deviation, $W_{t-1}$: Wealth Index for previous round}
\end{flushleft}
\end{table}

 The coefficient on lagged wealth ($\beta$) is highly significant and close to 0.54 for the older cohort and 0.58 for the younger cohort, suggesting that more than half of the marginal change in current economic status is explained by the previous round's status. These values indicate a relatively high level of economic immobility within households across rounds. The consistently small robust standard errors and extremely low $p$-values reinforce the statistical significance of the relationship. Intercept estimates are also stable across cohorts, reflecting a similar baseline level of economic well-being.

These findings underscore the importance of examining additional determinants such as health and education that may help explain the residual variation not captured by persistence alone.

\subsection{Influence of Health on Education}
\label{sec:results_health_edu}
 The cohort-round wise Durbin-Wu-Hausman test shows evidence for endogeneity (p-values $<0.05$) if the direct model \ref{equ:direc edu =health model} is applied. 
The two-stage least squares method is thus used.  
Since this approach takes into account the availability of previous round's subjective well-being level and IRT scores, the results are shown for Older Cohort's round 3 and 4 and for Younger Cohort's round 4 and 5. The sample sizes for older cohort's round 3 and 4 are 854 and 834 respectively while that for younger cohort's round 4 and 5 are 1792 and 1776 respectively.
Tables \ref{tab:results_first_stage} and \ref{tab:results_second_stage} show the results from the first and second stage respectively.

Table~\ref{tab:results_first_stage} reports the first-stage regression results. The dependent variable is the current round's subjective well-being, modeled as an ordinal outcome using a cumulative link model. Across all cohorts and rounds, lagged health and the wealth index (either current or previous or both) are found to be significant predictors of current health status. In particular, past subjective well-being shows consistently strong effects, suggesting temporal stability in health perceptions. 

\begin{table}[ht!]
\caption{Results from First Stage Model - Modeling Present Subjective Well-Being}
\label{tab:results_first_stage}
\centering
 \begin{subtable}[t]{\textwidth}
 \centering
  \caption{Older Cohort}
\begin{tabular}{lllllll}
\hline
\multicolumn{1}{c}{\multirow{2}{*}{\textbf{Term}}} & \multicolumn{3}{c}{\textbf{Round 3}}                                         & \multicolumn{3}{c}{\textbf{Round 4}}                                       \\ \cline{2-7} 
\multicolumn{1}{c}{}                                    & \multicolumn{1}{c}{\textbf{Estimate}} & \multicolumn{1}{c}{\textbf{SD}} & \multicolumn{1}{c}{\textbf{\textit{p-value}}} & \multicolumn{1}{c}{\textbf{ Estimate}} & \multicolumn{1}{c}{\textbf{SD}} & \multicolumn{1}{c}{\textbf{\textit{p-value}}} \\ \hline
$H_{t-1}=2$                                         & \multicolumn{1}{l}{0.28}     & \multicolumn{1}{l}{0.18}    &0.13           & \multicolumn{1}{l}{-0.08}         & 0.19 &0.66   \\ 
$H_{t-1}=3$  & \multicolumn{1}{l}{0.33}          & \multicolumn{1}{l}{0.18} &0.07    & \multicolumn{1}{l}{0.22}& 0.18&   0.23\\ 
$H_{t-1}=4$  & \multicolumn{1}{l}{0.45}           & \multicolumn{1}{l}{0.20} &0.03  & \multicolumn{1}{l}{0.55}          & 0.20&0.01              \\
$H_{t-1}=5$   & \multicolumn{1}{l}{0.59}          & \multicolumn{1}{l}{0.27}&0.03               & \multicolumn{1}{l}{0.58}           & 0.27&0.03              \\ 
$H_{t-1}=6$  & \multicolumn{1}{l}{0.88}          & \multicolumn{1}{l}{0.29}   &0            & \multicolumn{1}{l}{0.58}          & 0.29&0.05              \\ 
$W_t$ & \multicolumn{1}{l}{1.65}         & \multicolumn{1}{l}{0.63}    &0.01           & \multicolumn{1}{l}{1.05}          & 0.63&0.09               \\ 
$W_{t-1}$  & \multicolumn{1}{l}{1.10}         & \multicolumn{1}{l}{0.54}&0.04    & \multicolumn{1}{l}{1.26}         & 0.54&0.02             \\ \hline
\end{tabular}
\vspace{0.5cm}
\end{subtable}

 \begin{subtable}[t]{\textwidth}
 \centering
 \caption{Younger Cohort}
\begin{tabular}{lllllll}
\hline
\multicolumn{1}{c}{\multirow{2}{*}{\textbf{Term}}} & \multicolumn{3}{c}{\textbf{Round 4}}  & \multicolumn{3}{c}{\textbf{Round 5}}     \\ \cline{2-7} 
\multicolumn{1}{c}{}                                    & \multicolumn{1}{c}{\textbf{Estimate}} & \multicolumn{1}{c}{\textbf{SD}} & \multicolumn{1}{c}{\textbf{\textit{p-value}}} & \multicolumn{1}{c}{\textbf{ Estimate}} & \multicolumn{1}{c}{\textbf{SD}} & \multicolumn{1}{c}{\textbf{\textit{p-value}}} \\ \hline
$H_{t-1}=2$                                         & \multicolumn{1}{l}{0.09}     & \multicolumn{1}{l}{0.18}    &0.62           & \multicolumn{1}{l}{-0.56}         & 0.18 &0.00   \\ 
$H_{t-1}=3$  & \multicolumn{1}{l}{0.20}          & \multicolumn{1}{l}{0.18} &0.27 & \multicolumn{1}{l}{-0.26}& 0.18&   0.15\\ 
$H_{t-1}=4$  & \multicolumn{1}{l}{0.32}           & \multicolumn{1}{l}{0.20} &0.12  & \multicolumn{1}{l}{-0.01}          & 0.20&0.96             \\
$H_{t-1}=5$   & \multicolumn{1}{l}{0.37}          & \multicolumn{1}{l}{0.27}&0.18            & \multicolumn{1}{l}{0.15}           & 0.27&0.60              \\ 
$H_{t-1}=6$  & \multicolumn{1}{l}{0.48}          & \multicolumn{1}{l}{0.29}   &0.1            & \multicolumn{1}{l}{0.25}          & 0.29&0.39           \\ 
$W_t$ & \multicolumn{1}{l}{2.35}         & \multicolumn{1}{l}{0.63}    &0.00         & \multicolumn{1}{l}{1.54}          & 0.63&0.01               \\ 
$W_{t-1}$  & \multicolumn{1}{l}{1.17}         & \multicolumn{1}{l}{0.54}&0.03   & \multicolumn{1}{l}{1.43}         & 0.54&0.01          \\ \hline
\end{tabular}
\vspace{0.5em}\\
\begin{flushleft}
{\footnotesize $\dagger$ SD = Standard Deviation, $H_{t-1}$: Subjective well-being (Health) level for previous round, $W_t$: Wealth Index for current $t$, $W_{t-1}$: Wealth Index for previous round}
\end{flushleft}
\end{subtable}
\end{table}

Table~\ref{tab:results_second_stage} presents the second-stage OLS regression results, where the dependent variable is the current round's IRT-based math score, and the key regressor is the residualized health component ($H_r$) from the first-stage model. The coefficients on $H_r$ are generally small and statistically insignificant across most rounds and cohorts, indicating that after correcting for potential endogeneity, subjective health has a limited direct impact on educational achievement. In contrast, previous education ($E_{t-1}$) remains a strong and significant predictor in all models, confirming the persistence of cognitive performance over time. The wealth index also retains marginal significance in some specifications, suggesting it may influence educational outcomes through pathways beyond health.
\begin{table}[ht!]
\caption{Results from Second Stage Model : Modeling Present Maths(IRT) Scores}
\label{tab:results_second_stage}
\centering
 \begin{subtable}[t]{\textwidth}
 \centering
  \caption{Older Cohort}
\begin{tabular}{lllllll}
\hline
\multicolumn{1}{c}{\multirow{2}{*}{\textbf{Term}}} & \multicolumn{3}{c}{\textbf{Round 3}}                                         & \multicolumn{3}{c}{\textbf{Round 4}}                                       \\ \cline{2-7} 
\multicolumn{1}{c}{}                                    & \multicolumn{1}{c}{\textbf{Estimate}} & \multicolumn{1}{c}{\textbf{SD}} & \multicolumn{1}{c}{\textbf{\textit{p-value}}} & \multicolumn{1}{c}{\textbf{ Estimate}} & \multicolumn{1}{c}{\textbf{SD}} & \multicolumn{1}{c}{\textbf{\textit{p-value}}} \\ \hline
$H_r$&$-40.78$&35.07&0.24&10.26&34.12&0.76\\ 
$W_t$&80.39&25.97&0.002&18.35&25.53&0.47\\
$W_{t-1}$&32.15&22.94&0.16&43.76&20.80&0.03\\
$E_{t-1}$&0.46&0.02& $<10^{-16}$&0.82&0.03&$<10^{-16}$\\
$E_P$&1.29&0.40&0.0015&0.58&0.48&0.24\\ \hline
\end{tabular}
\vspace{0.5cm}
\end{subtable}

 \begin{subtable}[t]{\textwidth}
 \centering
 \caption{Younger Cohort}
\begin{tabular}{lllllll}
\hline
\multicolumn{1}{c}{\multirow{2}{*}{\textbf{Term}}} & \multicolumn{3}{c}{\textbf{Round 4}}                                         & \multicolumn{3}{c}{\textbf{Round 5}}                                       \\ \cline{2-7} 
\multicolumn{1}{c}{}                                    & \multicolumn{1}{c}{\textbf{Estimate}} & \multicolumn{1}{c}{\textbf{SD}} & \multicolumn{1}{c}{\textbf{\textit{p-value}}} & \multicolumn{1}{c}{\textbf{ Estimate}} & \multicolumn{1}{c}{\textbf{SD}} & \multicolumn{1}{c}{\textbf{\textit{p-value}}} \\ \hline
$H_r$&$0.86$&19.39&0.96&-9.70&15.54&0.53\\ 
$W_t$&49.17&15.50&0.001&40.45&15.46&0.008\\
$W_{t-1}$&43.69&14.13&0.002&43.49&14.29&0.002\\
$E_{t-1}$&0.55&0.02& $<10^{-16}$&0.58&0.02&$<10^{-16}$\\
$E_P$&1.27&0.40&0.0013&1.03&0.35&0.003\\ \hline
\end{tabular}
\vspace{0.5em}\\
\begin{flushleft}
{\footnotesize $\dagger$ SD = Robust Standard Deviation, $H_r$: Health residual from Stage 1, $W_t$: Wealth Index for current $t$, $W_{t-1}$: Wealth Index for previous round, $E_{t-1}$: Maths IRT scores for previous round, $E_P$: Mother's Education Level}
\end{flushleft}
\end{subtable}
\end{table}
\FloatBarrier

\subsection{Influence of Health and Education on Economic Well-being}
\label{sec:results_edu_wealth}
Table \ref{tab:results_cohort_persistence} presents the results from the cohort-round persistence of wealth index model. Table \ref{tab:results_wealth_residual_model} presents the results when the past subjective well-being values and past IRT scores are regressed on residual values from the persistence model.
\begin{table}[ht!]
\caption{Results from Cohort wise Persistence Model}
\label{tab:results_cohort_persistence}
\centering
 \begin{subtable}[t]{\textwidth}
 \centering
  \caption{Older Cohort}
\begin{tabular}{lllllll}
\hline
\multicolumn{1}{c}{\multirow{2}{*}{\textbf{Term}}} & \multicolumn{3}{c}{\textbf{Round 3}}                                         & \multicolumn{3}{c}{\textbf{Round 4}}                                       \\ \cline{2-7} 
\multicolumn{1}{c}{}                                    & \multicolumn{1}{c}{\textbf{Estimate}} & \multicolumn{1}{c}{\textbf{SD}} & \multicolumn{1}{c}{\textbf{\textit{p-value}}} & \multicolumn{1}{c}{\textbf{ Estimate}} & \multicolumn{1}{c}{\textbf{SD}} & \multicolumn{1}{c}{\textbf{\textit{p-value}}} \\ \hline
Intercept&$0.19$&0.01&$<10^{-16}$&0.27&0.01&$<10^{-16}$\\ 
$W_{t-1}$&0.7&0.02&$<10^{-16}$&0.63&0.02&$<10^{-16}$\\
\hline
\end{tabular}
\vspace{0.5cm}
\end{subtable}

 \begin{subtable}[t]{\textwidth}
 \centering
 \caption{Younger Cohort}
\begin{tabular}{lllllll}
\hline
\multicolumn{1}{c}{\multirow{2}{*}{\textbf{Term}}} & \multicolumn{3}{c}{\textbf{Round 4}}                                         & \multicolumn{3}{c}{\textbf{Round 5}}                                       \\ \cline{2-7} 
\multicolumn{1}{c}{}                                    & \multicolumn{1}{c}{\textbf{Estimate}} & \multicolumn{1}{c}{\textbf{SD}} & \multicolumn{1}{c}{\textbf{\textit{p-value}}} & \multicolumn{1}{c}{\textbf{ Estimate}} & \multicolumn{1}{c}{\textbf{SD}} & \multicolumn{1}{c}{\textbf{\textit{p-value}}} \\ \hline
Intercept&$0.22$&0.008&$<10^{-16}$&0.21&0.01&$<10^{-16}$\\ 
$W_{t-1}$&0.7&0.014&$<10^{-16}$&0.72&0.015&$<10^{-16}$\\
\hline
\end{tabular}
\vspace{0.5em}\\
\begin{flushleft}
{\footnotesize $\dagger$ SD = Robust Standard Deviation, $W_{t-1}$: Wealth Index for previous round}
\end{flushleft}
\end{subtable}
\end{table}

\begin{table}[ht!]
\caption{Results from Regressing Past Health and Education on Residuals from Persistence Model (table \ref{tab:results_cohort_persistence})}
\label{tab:results_wealth_residual_model}
\centering
 \begin{subtable}[t]{\textwidth}
 \centering
  \caption{Older Cohort}
\begin{tabular}{lllllll}
\hline
\multicolumn{1}{c}{\multirow{2}{*}{\textbf{Term}}} & \multicolumn{3}{c}{\textbf{Round 3}}                                         & \multicolumn{3}{c}{\textbf{Round 4}}                                       \\ \cline{2-7} 
\multicolumn{1}{c}{}                                    & \multicolumn{1}{c}{\textbf{Estimate}} & \multicolumn{1}{c}{\textbf{SD}} & \multicolumn{1}{c}{\textbf{\textit{p-value}}} & \multicolumn{1}{c}{\textbf{ Estimate}} & \multicolumn{1}{c}{\textbf{SD}} & \multicolumn{1}{c}{\textbf{\textit{p-value}}} \\ \hline
$H_{t-1}=2$ &   0.015 &0.01&0.14 & -0.05&0.02&0.007\\ 
$H_{t-1}=3$ &0.012&0.01&0.25&-0.04&0.02&0.03\\ 
$H_{t-1}=4$  &0.03&0.01&0.002 &-0.03&0.02&0.054\\
$H_{t-1}=5$ &0.03&0.01&0.047 &-0.03&0.02&0.067\\ 
$H_{t-1}=6$  & 0.02&0.02&0.12&-0.04&0.02&0.026\\ 
$E_{t-1}$  & 0 & $<10^{-5}$&0.14 &0.0001& $<10^{-16}$& 0.005 \\ \hline
\end{tabular}
\vspace{0.5cm}
\end{subtable}

\begin{subtable}[t]{\textwidth}
 \centering
  \caption{Younger Cohort}
\begin{tabular}{lllllll}
\hline
\multicolumn{1}{c}{\multirow{2}{*}{\textbf{Term}}} & \multicolumn{3}{c}{\textbf{Round 4}}                                         & \multicolumn{3}{c}{\textbf{Round 5}}    \\ \cline{2-7} 
\multicolumn{1}{c}{}                                    & \multicolumn{1}{c}{\textbf{Estimate}} & \multicolumn{1}{c}{\textbf{SD}} & \multicolumn{1}{c}{\textbf{\textit{p-value}}} & \multicolumn{1}{c}{\textbf{ Estimate}} & \multicolumn{1}{c}{\textbf{SD}} & \multicolumn{1}{c}{\textbf{\textit{p-value}}} \\ \hline
$H_{t-1}=2$ & -0.021 &0.01&0.04 & 0.04&0.01&0.005\\ 
$H_{t-1}=3$ &-0.017&0.009&0.07&0.04&0.01&0.002\\ 
$H_{t-1}=4$  &-0.005&0.009&0.072 &0.05&0.01&0.0007\\
$H_{t-1}=5$ &0.003&0.01&0.73 &0.04&0.01&0.003\\ 
$H_{t-1}=6$  &-0.006&0.009&0.44&0.05&0.01&0.0001\\ 
$E_{t-1}$  & 0.0002 & 0.00003&$<10^{-8}$ &0.0001& 0.00002& $<10^{-5}$ \\ \hline
\end{tabular}
\vspace{0.5em}\\
\begin{flushleft}
{\footnotesize $\dagger$ SD = Robust Standard Deviation, $H_{t-1}$: Subjective well-being (Health) level for previous round, $E_{t-1}$: Maths IRT scores for previous round}
\end{flushleft}
\end{subtable}

\end{table}

Across cohorts and rounds, the coefficients for previous education ($E_{t-1}$) are consistently positive, indicating that higher prior educational attainment is associated with improved household economic standing beyond what is captured by wealth persistence. Although the magnitudes are modest, the positive direction is stable across specifications, and coefficients are statistically significant for the younger cohort and marginally for the older cohort's round 4. This reflects education’s reinforcing effect on economic opportunity, even in older groups where transitions into adulthood may already be well underway.

In contrast, the coefficients on health categories show mixed signs. This may suggest non-linear or threshold effects of self-perceived health that are not fully captured by the ordinal scale, or it may reflect greater measurement noise in self-assessed well-being. Moreover, the health coefficients are generally not statistically significant, limiting the confidence in any particular directional interpretation.

Overall, the signs and magnitudes in Table~\ref{tab:results_wealth_residual_model} underscore that prior cognitive performance (as measured through IRT scores) is a more robust and reliable predictor of subsequent economic status than self-reported health, particularly in this cohort.

\begin{table}
\caption{Adjusted $R^2$ for both the Models}
\label{tab: results_ Adjusted R^2}
\centering
\begin{tabular}{llll}  
\hline
\textbf{Cohort} &\textbf{ Round }& \textbf{Persistence Model $R^2$} & \textbf{Model \ref{equ:past_health_edu_wealth} $R^2$} \\ \hline
Older & 3 & 0.523 & 0.014 \\ 
Older & 4 & 0.658 & 0.007 \\ 
Younger & 4 & 0.571 & 0.187 \\ 
Younger & 5 & 0.581 & 0.019 \\ \hline
\end{tabular}
\end{table}

Table \ref{tab: results_ Adjusted R^2} shows that the predictive strength of past Wealth Index is notably high across all cohorts and rounds, with the persistence model explaining between 52\% and 66\% of the variance in current Wealth. In contrast, the residual-based Model \ref{equ:past_health_edu_wealth}, which tests whether Health and Education explain variation not captured by past Wealth, yields much lower \( R^2 \) values. For the older cohort, these variables account for less than 2\% of the residual variance in both rounds. However, for the younger cohort in Round 4, Health and Education explain a relatively higher 18.7\% of the residual variance, suggesting their growing importance at earlier life stages. Overall, while past Wealth dominates in predicting current economic well-being, the added influence of Health and Education appears more cohort- and age-dependent.

The stronger and more consistent significance of education-related predictors in the younger cohort suggests a growing role of cognitive achievement in influencing economic well-being, potentially reflecting structural changes in opportunity dynamics over time. Across models, both past education (as measured by IRT scores) and subjective well-being contribute meaningfully to explaining current wealth, beyond what is captured by past wealth alone. 

The significance of different health categories across rounds and cohorts points to nuanced, cohort-specific effects that merit further exploration. Nonetheless, education emerges as a consistently significant predictor, particularly in the younger cohort, where schooling policies and cognitive development programs may have had a more pronounced impact.

Overall, these findings affirm that both health and education contribute to shaping economic well-being, complementing the strong role played by past wealth. The persistence of education’s influence across rounds underscores its central role in promoting upward mobility, especially in low-resource settings such as India.

\subsection{Causal Graphs}
\label{sec:results_causal_graphs}

This section synthesizes the regression-based findings into directed causal graphs, providing a visual summary of statistically significant relationships among Health (H), Education (E), and Economic Well-being (Wealth, W) across cohorts and survey rounds.

\begin{figure}[ht!]
    \centering
    \begin{subfigure}{0.45\textwidth}
\centering
 \includegraphics[width=\linewidth]{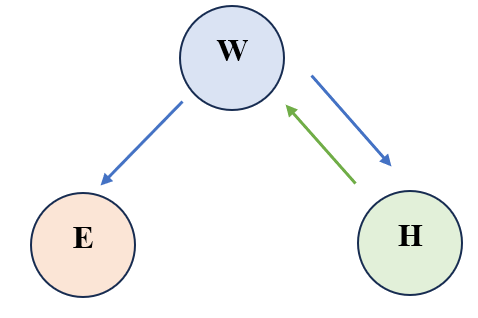}
\caption{Older Cohort - Round 3}
    \end{subfigure}
  \hspace{0.5em} 
    \begin{subfigure}{0.45\textwidth}
        \centering
        \includegraphics[width=\linewidth]{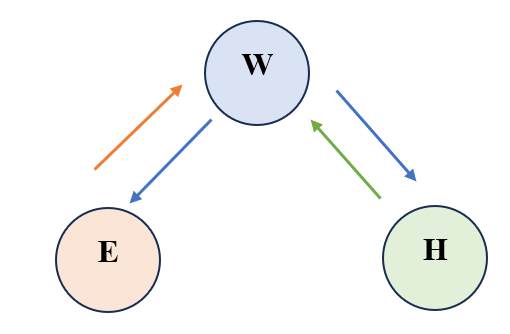}
     \caption{Older Cohort - Round 4}
    \end{subfigure}

    \vspace{0.5em} 
    \begin{subfigure}{0.45\textwidth}
        \centering
        \includegraphics[width=\linewidth]{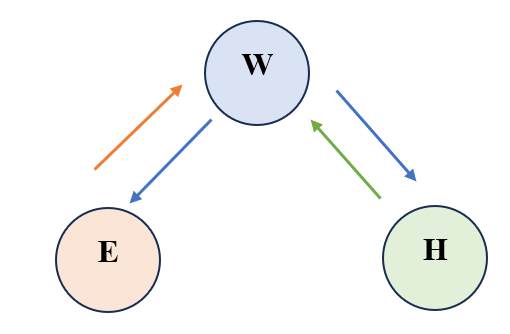}
        \caption{Younger Cohort - Round 4}
    \end{subfigure}
    \hspace{0.5em} 
    \begin{subfigure}{0.45\textwidth}
\centering
\includegraphics[width=\linewidth]{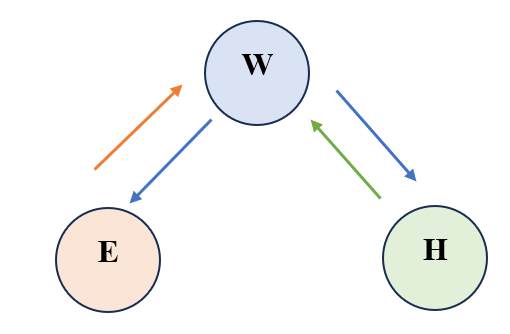}
\caption{Younger Cohort - Round 5}
    \end{subfigure}
 \caption{Causal Graphs from Model Results}
 \label{CausalGraphs}
\end{figure}

Figure~\ref{CausalGraphs} presents these causal diagrams for both the Older and Younger cohorts. Each arrow represents a statistically significant directional association, as inferred from model estimates. The graph construction follows the logic outlined below:

\begin{enumerate}[label=(\alph*)]
    \item \textbf{H $\rightarrow$ E:} This edge is determined based on the results from Table~\ref{tab:results_second_stage}, where the residualized health term ($H_r$) was included as a regressor in the second stage of the model for education. As $H_r$ was not statistically significant for either cohort, no arrow from H to E is included.
    
    \item \textbf{E $\rightarrow$ H:} This direction is not evaluated in this study, as our framework focuses on early-life influences of health and education on future outcomes. Given the age range of participants, reverse causality from education to health is not considered plausible and is therefore omitted from the graph.

    \item \textbf{W $\rightarrow$ H and W $\rightarrow$ E:} These edges are included if either the current or previous wealth index ($W_t$ or $W_{t-1}$) was statistically significant in the first-stage health model (Table~\ref{tab:results_first_stage}) or the education model (Table~\ref{tab:results_second_stage}), respectively.

    \item \textbf{E $\rightarrow$ W:} An edge from education to wealth is included if the previous IRT score ($E_{t-1}$) is statistically significant in the wealth residual regression model (Table~\ref{tab:results_wealth_residual_model}).

    \item \textbf{H $\rightarrow$ W:} A directional edge from health to wealth is included if at least one level of the categorical health variable ($H_{t-1}$) is statistically significant in Table~\ref{tab:results_wealth_residual_model}.
\end{enumerate}
These causal diagrams serve to concisely represent cohort-specific and round-specific differences in the strength and direction of relationships, and they guide interpretation of the multivariate dependencies across the study variables.
The arrow from Wealth to Health and Wealth to Education is consistently present, reinforcing the idea that better economic conditions support improved subjective well-being and educational opportunities. This may reflect access to nutrition, healthcare, or reduced psychosocial stress—all critical determinants of child health.

The influence of Education on Wealth (E $\rightarrow$ W) is present in three of the four graphs, and notably stronger in the younger cohort. This suggests that cognitive achievement is increasingly translating into economic advantages, possibly due to better integration of education with labor market opportunities in later survey rounds. It also reflects that education is becoming a more central pathway for upward mobility among younger individuals.

Interestingly, there is no arrow from Health to Education (H $\rightarrow$ E), indicating that—after controlling for wealth and other covariates—subjective health does not exert a statistically significant direct influence on cognitive outcomes in this dataset. This could be due to the self-reported nature of the health measure or potential mediation by wealth.

In the younger cohorts and older cohort- round 4, the bidirectional association between Wealth and Education is visually prominent (arrows in both directions), indicating the reinforcing feedback loop: better education improves economic status, and improved economic status, in turn, enhances educational opportunity. This feedback dynamic may reflect structural shifts in access to opportunity or recent policy interventions.

Overall, the causal graphs highlight education as a more consistent and robust lever for improving long-term economic well-being, while health appears to operate more indirectly through wealth. The evolving patterns across rounds and cohorts also underscore the importance of age, timing, and context in shaping developmental trajectories.
\section{Concluding Remarks}
\label{sec:conclusion}
This study explores the dynamic and potentially causal relationships among childhood health, cognitive ability, and economic well-being in the Indian context using longitudinal data from the Young Lives Survey. Through a combination of regression models, instrumental variable techniques, and causal graph analysis, the results offer nuanced evidence on how early-life conditions shape long-term developmental outcomes.

This study yields three principal findings regarding the intergenerational dynamics of health, education, and economic well-being in India:

\begin{enumerate}[label=(\alph*)]
    \item \textbf{High persistence in economic well-being:} Household wealth exhibits strong temporal stability across survey rounds and cohorts, suggesting structural inertia and limited short-term mobility in economic status.

    \item \textbf{Wealth as a foundational driver:} Current and past wealth consistently influence both educational attainment and self-reported health outcomes, underscoring the critical role of household resources in shaping early-life development.

    \item \textbf{Education as the strongest forward predictor:} Cognitive achievement, measured via IRT-based mathematics scores emerges as the most robust and consistent predictor of future economic well-being, particularly among the younger cohort. While subjective health contributes to economic outcomes, its direct influence on educational attainment is weaker and non-significant.

\end{enumerate}

Despite these insights, several limitations should be acknowledged.
First, the generalizability of the findings is constrained by the geographic scope of the Young Lives Survey in India, which covers children from Andhra Pradesh and Telangana. These states may differ from others in terms of socio-economic context, public service access, and cultural norms. Thus, caution is warranted when extending the conclusions to the broader Indian population.
Second, childhood health is proxied using self-reported subjective well-being on a 9-point ordinal scale. Although subjective health measures are widely used and correlate with objective outcomes, they may be prone to reporting bias, particularly among younger respondents or across socio-economic groups with differing perceptions of health.
Third, economic well-being is measured using a composite wealth index based on household assets and amenities, which is a commonly accepted proxy in the absence of reliable income data. However, this index may not capture short-term income fluctuations, informal labor earnings, or intra-household disparities. As such, it reflects long-run household conditions rather than transient economic shocks.

Acknowledging these limitations provides a foundation for future work that builds on the present findings with expanded data, refined measures, and alternative identification strategies.

Overall, the findings emphasize the enduring significance of wealth and education in shaping intergenerational mobility, while suggesting that investments in learning outcomes may offer the most reliable path to upward economic transition in resource-constrained settings like India.
\section{Discussion}
\label{sec:discussion}
The results of this study carry important implications for both policy and future research. From a policy perspective, the consistent influence of household wealth on children’s health and education underscores the foundational role of economic conditions in shaping early-life development. This suggests a need for targeted support to low-income families, such as conditional cash transfers, nutritional subsidies, or asset-building programs, to reduce structural barriers that constrain human capital formation.

More critically, the robust predictive power of cognitive ability, particularly mathematics performance measured via IRT scores points to the rising importance of foundational education in driving economic mobility. This lends empirical support to India’s National Education Policy \citep{NEP2020}, which emphasizes numeracy, reasoning skills, and equitable access to quality education. Investments in early-grade learning are likely to yield long-term returns in economic well-being, especially for the most vulnerable populations.

At the same time, the relatively weaker and inconsistent influence of childhood health, after adjusting for wealth and other covariates suggests that standalone health interventions may not be sufficient to produce economic gains unless they are integrated with education-focused strategies. Multidimensional programs such as school-based nutrition, free health check-ups, and learning incentives are more likely to break cycles of disadvantage than siloed approaches.

The observed cohort differences also reveal that timing matters: the impact of education on wealth appears stronger in the younger cohort, suggesting that intervening earlier in the life course, particularly during primary school years may maximize the returns to both health and educational investments. This insight can guide the targeting and sequencing of policy interventions aimed at promoting upward mobility in resource-constrained settings.

For future work, incorporating additional dimensions such as gender, caste, or rural-urban disparities may offer a more complete picture of intergenerational inequality. Furthermore, integrating Round 6 of the Young Lives dataset will allow examination of post-COVID dynamics and labor market outcomes in early adulthood. Finally, replicating this analysis in other low- and middle-income contexts would help uncover whether the observed patterns generalize or vary across institutional settings.

This study responds to the need, outlined in previous research \citep[e.g.,][]{Currie_Vogal_2013, Bellani2018}, for integrated empirical frameworks that move beyond isolated effects of health or education to examine how these domains interact over time to shape economic trajectories. While prior studies in high-income settings have highlighted the importance of education as a mediator of early disadvantage, there has been limited causal evidence from developing country contexts, particularly using longitudinal data. By leveraging panel data from the Young Lives Survey, this study provides empirical insight into the dynamics of early-life inequality in India, highlighting that while household wealth underpins access to both health and education, it is cognitive achievement that appears most predictive of future economic well-being.

Overall, this study reinforces the need for coordinated, early-life policy interventions that jointly address health, education, and economic disadvantage. In doing so, it contributes to the growing literature that uses statistical modeling and causal inference to inform evidence-based development policy.

\section*{Data Disclosure Statement}
The data used in this publication come from Young Lives, a 20-year study of childhood poverty and transitions to adulthood in Ethiopia, India, Peru and Vietnam (\url{www.younglives.org.uk}). Young Lives is funded by UK aid from the Foreign, Commonwealth \& Development Office and a number of further funders. The views expressed here are those of the authors. They are not necessarily those of Young Lives, the University of Oxford, FCDO or other funders.

\end{document}